\documentclass{article}
\usepackage{kr}

\usepackage{times}
\usepackage{soul}
\usepackage{url}
\usepackage[hidelinks]{hyperref}
\usepackage[utf8]{inputenc}
\usepackage[small,figurename=Listing]{caption} 
\usepackage{graphicx}
\usepackage{amsmath}
\usepackage{amsthm}
\usepackage{booktabs}
\usepackage{algorithm}
\usepackage{algorithmic}
\newcommand{\basicDatastructuresRepoUrl}{https://github.com/monsterkrampe/Basic-Lean-Datastructures/tree/v0.1.0}
\newcommand{\basicDatastructuresRepoLinkBase}[2]{\href{\basicDatastructuresRepoUrl/#1}{\textcolor{leansymbolcolor}{\texttt{#2}}}}
\newcommand{\basicDatastructuresRepoLinkCode}[2]{\basicDatastructuresRepoLinkBase{BasicLeanDatastructures/#1}{#2}}

\newcommand{\treeRepoUrl}{https://github.com/monsterkrampe/Possibly-Infinite-Trees/tree/v0.1.0}
\newcommand{\treeRepoLinkBase}[2]{\href{\treeRepoUrl/#1}{\textcolor{leansymbolcolor}{\texttt{#2}}}}
\newcommand{\treeRepoLinkCode}[2]{\treeRepoLinkBase{PossiblyInfiniteTrees/#1}{#2}}

\newcommand{\ruleRepoUrl}{https://github.com/monsterkrampe/Existential-Rules-in-Lean/tree/v0.2.0}
\newcommand{\ruleRepoLinkBase}[2]{\href{\ruleRepoUrl/#1}{\textcolor{leansymbolcolor}{\texttt{#2}}}}
\newcommand{\ruleRepoLinkCode}[2]{\ruleRepoLinkBase{ExistentialRules/#1}{#2}}

\newcommand{\Set}{\basicDatastructuresRepoLinkCode{Set/Basic.lean\#L15}{Set}}
\newcommand{\SetFinite}{\basicDatastructuresRepoLinkCode{Set/Finite.lean\#L25}{finite}}

\newcommand{\InfiniteList}{\treeRepoLinkCode{PossiblyInfiniteList/InfiniteList.lean\#L27}{InfiniteList}}
\newcommand{\PossiblyInfiniteList}{\treeRepoLinkCode{PossiblyInfiniteList/PossiblyInfiniteList.lean\#L30}{PossiblyInfiniteList}}

\newcommand{\ListHead}{\treeRepoLinkCode{PossiblyInfiniteList/PossiblyInfiniteList.lean\#L55}{head}}

\newcommand{\ListTail}{\treeRepoLinkCode{PossiblyInfiniteList/PossiblyInfiniteList.lean\#L58}{tail}}
\newcommand{\ListMemRec}{\treeRepoLinkCode{PossiblyInfiniteList/PossiblyInfiniteList.lean\#L218}{mem\_rec}}
\newcommand{\Generate}{\treeRepoLinkCode{PossiblyInfiniteList/PossiblyInfiniteList.lean\#L313}{generate}}

\newcommand{\FiniteDegreeTree}{\treeRepoLinkCode{PossiblyInfiniteTree/FiniteDegreeTree/Basic.lean\#L35}{FiniteDegreeTree}}
\newcommand{\GenerateBranch}{\treeRepoLinkCode{PossiblyInfiniteTree/FiniteDegreeTree/Basic.lean\#L640}{generate\_branch}}

\newcommand{\Signature}{\ruleRepoLinkCode{Terms/Basic.lean\#L23}{Signature}}
\newcommand{\SkolemFS}{\ruleRepoLinkCode{Terms/Basic.lean\#L43}{SkolemFS}}

\newcommand{\GroundTerm}{\ruleRepoLinkCode{Terms/Basic.lean\#L180}{GroundTerm}}
\newcommand{\SkolemTerm}{\ruleRepoLinkCode{Terms/Basic.lean\#L51}{SkolemTerm}}
\newcommand{\VarOrConst}{\ruleRepoLinkCode{Terms/Basic.lean\#L83}{VarOrConst}}

\newcommand{\cyclicOfDepthTooBig}{\ruleRepoLinkCode{Terms/Cyclic.lean\#L126}{cyclic\_of\_depth\_too\_big}}

\newcommand{\GeneralizedAtom}{\ruleRepoLinkCode{AtomsAndFacts/Basic.lean\#L29}{GeneralizedAtom}}
\newcommand{\Atom}{\ruleRepoLinkCode{AtomsAndFacts/Basic.lean\#L40}{Atom}}
\newcommand{\FunctionFreeAtom}{\ruleRepoLinkCode{AtomsAndFacts/Basic.lean\#L49}{FunctionFreeAtom}}
\newcommand{\FunctionFreeConjunction}{\ruleRepoLinkCode{AtomsAndFacts/Basic.lean\#L107}{FunctionFreeConjunction}}
\newcommand{\Rule}{\ruleRepoLinkCode{AtomsAndFacts/Basic.lean\#L162}{Rule}}

\newcommand{\RuleSet}{\ruleRepoLinkCode{AtomsAndFacts/Basic.lean\#L263}{RuleSet}}
\newcommand{\Fact}{\ruleRepoLinkCode{AtomsAndFacts/Basic.lean\#L408}{Fact}}
\newcommand{\FunctionFreeFact}{\ruleRepoLinkCode{AtomsAndFacts/Basic.lean\#L411}{FunctionFreeFact}}
\newcommand{\FactSet}{\ruleRepoLinkCode{AtomsAndFacts/Basic.lean\#L501}{FactSet}}
\newcommand{\Database}{\ruleRepoLinkCode{AtomsAndFacts/Basic.lean\#L681}{Database}}
\newcommand{\KnowledgeBase}{\ruleRepoLinkCode{AtomsAndFacts/Basic.lean\#L774}{KnowledgeBase}}

\newcommand{\GroundSubstitution}{\ruleRepoLinkCode{AtomsAndFacts/SubstitutionsAndHomomorphisms.lean\#L175}{GroundSubstitution}}

\newcommand{\GroundTermMapping}{\ruleRepoLinkCode{AtomsAndFacts/SubstitutionsAndHomomorphisms.lean\#L233}{GroundTermMapping}}

\newcommand{\PreTrigger}{\ruleRepoLinkCode{Triggers/Basic.lean\#L27}{PreTrigger}}

\newcommand{\loaded}{\ruleRepoLinkCode{Triggers/Basic.lean\#L494}{loaded}}

\newcommand{\PreTriggerEquiv}{\ruleRepoLinkCode{Triggers/Basic.lean\#L541}{equiv}}
\newcommand{\LaxObsolescenceCondition}{\ruleRepoLinkCode{Triggers/Obsolescence.lean\#L36}{LaxObsolescenceCondition}}
\newcommand{\ObsolescenceCondition}{\ruleRepoLinkCode{Triggers/Obsolescence.lean\#L41}{ObsolescenceCondition}}
\newcommand{\SkolemObsolescence}{\ruleRepoLinkCode{Triggers/Obsolescence.lean\#L61}{SkolemObsolescence}}
\newcommand{\RestrictedObsolescence}{\ruleRepoLinkCode{Triggers/Obsolescence.lean\#L96}{RestrictedObsolescence}}
\newcommand{\Trigger}{\ruleRepoLinkCode{Triggers/Obsolescence.lean\#L136}{Trigger}}
\newcommand{\TriggerActive}{\ruleRepoLinkCode{Triggers/Obsolescence.lean\#L147}{active}}
\newcommand{\RTrigger}{\ruleRepoLinkCode{Triggers/RTrigger.lean\#L25}{RTrigger}}

\newcommand{\ChaseNode}{\ruleRepoLinkCode{ChaseSequence/ChaseNode.lean\#L23}{ChaseNode}}

\newcommand{\ChaseDerivationSkeleton}{\ruleRepoLinkCode{ChaseSequence/ChaseDerivationSkeleton.lean\#L56}{ChaseDerivationSkeleton}}
\newcommand{\ChaseDerivationHead}{\ruleRepoLinkCode{ChaseSequence/ChaseDerivationSkeleton.lean\#L103}{head}}
\newcommand{\ChaseDerivationNext}{\ruleRepoLinkCode{ChaseSequence/ChaseDerivationSkeleton.lean\#L119}{next}}

\newcommand{\ChaseDerivationPrec}{\ruleRepoLinkCode{ChaseSequence/ChaseDerivationSkeleton.lean\#L378}{predecessor}}

\newcommand{\ChaseDerivation}{\ruleRepoLinkCode{ChaseSequence/ChaseDerivation.lean\#L44}{ChaseDerivation}}
\newcommand{\ChaseDerivationSuffix}{\ruleRepoLinkCode{ChaseSequence/ChaseDerivation.lean\#L174}{IsSuffix}}
\newcommand{\ChaseDerivationFairnessPrime}{\ruleRepoLinkCode{ChaseSequence/ChaseDerivation.lean\#L189}{fairness'}}
\newcommand{\ChaseDerivationTail}{\ruleRepoLinkCode{ChaseSequence/ChaseDerivation.lean\#L215}{tail}}
\newcommand{\ChaseDerivationMemRec}{\ruleRepoLinkCode{ChaseSequence/ChaseDerivation.lean\#L259}{mem\_rec}}
\newcommand{\ChaseDerivationFairnessPrec}{\ruleRepoLinkCode{ChaseSequence/ChaseDerivation.lean\#L359}{fairness\_prec}}

\newcommand{\ChaseBranch}{\ruleRepoLinkCode{ChaseSequence/ChaseBranch.lean\#L27}{ChaseBranch}}
\newcommand{\ChaseBranchResultModelsKb}{\ruleRepoLinkCode{ChaseSequence/ChaseBranch.lean\#L68}{result\_models\_kb}}

\newcommand{\TreeDerivation}{\ruleRepoLinkCode{ChaseSequence/TreeDerivation.lean\#L80}{TreeDerivation}}

\newcommand{\TreeDerivationGenerateSubderivation}{\ruleRepoLinkCode{ChaseSequence/TreeDerivation.lean\#L1136}{generate\_subderivation}}
\newcommand{\TreeDerivationGenerateSubderivationMemBranches}{\ruleRepoLinkCode{ChaseSequence/TreeDerivation.lean\#L1144}{generate\_subderivation\_mem\_branches}}

\newcommand{\ChaseTree}{\ruleRepoLinkCode{ChaseSequence/ChaseTree.lean\#L27}{ChaseTree}}
\newcommand{\ChaseTreeResultModelsKb}{\ruleRepoLinkCode{ChaseSequence/ChaseTree.lean\#L93}{result\_models\_kb}}

\newcommand{\InductiveHomomorphismResult}{\ruleRepoLinkCode{ChaseSequence/Universality.lean\#L37}{InductiveHomomorphismResult}}
\newcommand{\HomStep}{\ruleRepoLinkCode{ChaseSequence/Universality.lean\#L209}{hom\_step}}
\newcommand{\ChaseTreeResultIsUniversal}{\ruleRepoLinkCode{ChaseSequence/Universality.lean\#L272}{chaseTreeResultIsUniversal}}

\newcommand{\DeterministicChaseBranchResultUniversallyModelsKb}{\ruleRepoLinkCode{ChaseSequence/Deterministic.lean\#L412}{detCBResultUniversallyModelsKb}}

\newcommand{\homForNodeExtendableToResult}{\ruleRepoLinkCode{AlternativeMatches/HomomorphismExtension.lean\#L228}{hom\_for\_node\_extendable\_to\_result}}

\newcommand{\resultIsWeakCoreOfNoAltMatch}{\ruleRepoLinkCode{AlternativeMatches/Chase.lean\#L65}{result\_isWeakCore\_of\_noAltMatch}}
\newcommand{\resultIsStrongCoreOfNoAltMatch}{\ruleRepoLinkCode{AlternativeMatches/Chase.lean\#L598}{result\_isStrongCore\_of\_noAltMatch}}

\newcommand{\parallelDeterminizedChase}{\ruleRepoLinkCode{ChaseSequence/Termination/ParallelDeterminizedChase.lean\#L357}{parallelDeterminizedChase}}
\newcommand{\MfaObsolescenceCondition}{\ruleRepoLinkCode{ChaseSequence/Termination/MfaLike.lean\#L134}{MfaObsolescenceCondition}}
\newcommand{\blocksObs}{\ruleRepoLinkCode{ChaseSequence/Termination/MfaLike.lean\#L146}{blocks\_obs}}
\newcommand{\DeterministicSkolemObsolescence}{\ruleRepoLinkCode{ChaseSequence/Termination/MfaLike.lean\#L153}{DeterministicSkolemObsolescence}}
\newcommand{\BlockingObsolescence}{\ruleRepoLinkCode{ChaseSequence/Termination/MfaLike.lean\#L231}{BlockingObsolescence}}
\newcommand{\criticalInstance}{\ruleRepoLinkCode{ChaseSequence/Termination/MfaLike.lean\#L636}{criticalInstance}}
\newcommand{\mfaKb}{\ruleRepoLinkCode{ChaseSequence/Termination/MfaLike.lean\#L690}{mfaKb}}
\newcommand{\mfaSet}{\ruleRepoLinkCode{ChaseSequence/Termination/MfaLike.lean\#L708}{mfaSet}}
\newcommand{\TerminatesOfMfaSetFinite}{\ruleRepoLinkCode{ChaseSequence/Termination/MfaLike.lean\#L983}{terminates\_of\_mfaSet\_finite}}
\newcommand{\isMfa}{\ruleRepoLinkCode{ChaseSequence/Termination/MfaLike.lean\#L1101}{isMfa}}
\newcommand{\TerminatesOfIsMfa}{\ruleRepoLinkCode{ChaseSequence/Termination/MfaLike.lean\#L1105}{terminates\_of\_isMfa}}

\usepackage{listings}
\usepackage{todonotes}

\usepackage{color}
\definecolor{keywordcolor}{rgb}{0.7, 0.1, 0.1}   
\definecolor{tacticcolor}{rgb}{0.0, 0.1, 0.6}    
\definecolor{commentcolor}{rgb}{0.4, 0.4, 0.4}   
\definecolor{symbolcolor}{rgb}{0.0, 0.1, 0.6}    
\definecolor{sortcolor}{rgb}{0.1, 0.5, 0.1}      
\definecolor{attributecolor}{rgb}{0.7, 0.1, 0.1} 

\definecolor{leansymbolcolor}{RGB}{32,16,160}

\newtheorem{example}{Example}

\newtheorem{definition}{Definition}
\newtheorem{remark}{Remark}

\title{The Chase in Lean - Crafting a Formal Library for Existential Rule Research}

\author{%
    Lukas Gerlach
    \affiliations
    Knowledge-Based Systems Group, TU Dresden, Germany
    \emails
    lukas.gerlach@tu-dresden.de    
}

\begin{document}

\maketitle

\begin{abstract}
  The chase is a sound, complete, but possibly non-terminating algorithm for reasoning with existential rules (aka. tuple-generating dependencies), 
  a highly expressive knowledge representation language.
  Although the procedure appears simple, research on theoretical properties and optimization for practical implementations has grown 
  to a point where verifying correctness and reproducing proofs becomes challenging and intuition can sometimes be misleading.
  Lean is a purely functional programming language and interactive theorem prover 
  whose community actively develops formal libraries for mathematics (Mathlib) and computer science (CSLib).
  In this work, we present our own endeavor of crafting a Lean framework around existential rules and the chase.
  We discuss design decisions concerning the nuances of chase definitions commonly found in the literature and show how these translate into Lean.
  To illustrate the framework's capabilities using known results, we show that the result of a chase is a universal model
  and outline the formalization for proving that without so-called ``alternative matches'' it is even a core.
  Beyond existing literature, we unify sufficient chase termination conditions in the likeness of Model-Faithful Acyclicity (MFA) 
  into a common framework while also adding support for constants in rules.
\end{abstract}

\section{Introduction}

The chase \cite{Chase} is a common reasoning algorithm in knowledge representation and reasoning. 
It computes a bottom-up materialization for a given knowledge base 
consisting of a set of existential rules \cite{ExistentialRules} and a database.
After an introductory example,
we discuss properties of the chase that we pick as candidates for a machine-verifiable formalization,
motivate the need for formal verification in this setting, and 
give an overview of our specific contributions made in building a Lean library.

\begin{example}\label{exp:mainExample}
  Consider the database $D = \{P(a, b)\}$ and the singleton rule set $R$:
  $$P(x, y) \to S(y) \lor \exists z. P(y, z) \land P(z, y)$$
  A chase of $\langle R, D \rangle$ is a possibly infinite tree of fact sets that starts on the root $D$ and tries to ``apply'' a rule to each node to obtain its children, one for each disjunct. 
  Depending on the variant of chase algorithm, there are different conditions for when a node receives children.
  For example, in the restricted (aka. standard) chase,
  if all rules are already satisfied in a node $n$, then $n$ has no children.
  
  In this example, 
  the root is $\{ P(a, b) \}$ and has two children 
  $\{ P(a,b), S(b) \}$ and $\{ P(a, b), \allowbreak P(b, f^{1,1}_z(b)), \allowbreak P(f^{1,1}_z(b), b) \}$. 
  Note that both of these children satisfy the single rule in $R$, and therefore they are the leaves of the tree.
\end{example}

We use Skolem terms ($f^{1,1}_z(b)$) as a naming convention for ``fresh'' terms 
introduced for existential variables \cite{SkolemChase}. 
The function symbol $f^{1,1}_z$ uniquely identifies the rule id ($1$), the (zero-based) disjunct index ($1$) and the variable ($z$).
The arguments are the terms for the variables that occur on both sides of the implication (i.e. $y \mapsto b$).

\subsection{Interesting Properties of the Chase}

The result of the chase, i.e. the set of all fact sets that can be built by taking the union of all fact sets along a branch of the chase tree, is a \emph{universal model set} \cite{UniversalModelSet}. 
In the absence of disjunctions, the result is a single fact set, which is a universal model \cite{ChaseRevisited}. 
Universal model sets are of interest for conjunctive query answering~\cite{Alice}.

Beyond universality, it is desirable to create the smallest possible models, called ``cores''.
While the core chase \cite{ChaseRevisited} is able to achieve this in theory, 
it does so by finding homomorphisms that allow collapsing fact sets into substructures, which is infeasible in practice.
Fairly recent work coins the term ``alternative match'' for redundancies introduced during the chase.
Avoiding alternative matches directly yields a core without the need for any additional computations \cite{RestrictedChaseCores}. 

With these nice properties, there is a catch: chase termination.
For example, for the rule $R(x,y) \to \exists z. R(y,z)$ and the database $\{R(a,b)\}$, 
the chase will produce an infinite branch. 
In the general case, we cannot decide if the chase terminates for a given set of rules \cite{ChaseTerminationUndecidable,AnatomyOfTheChase}.
There is a whole line of research defining classes of rule sets with guaranteed chase termination \cite{WeakAcyclicity,JointAcyclicity,ExtendingAcyclicityNotionsForExistentialRules,KarimiTermination,PhilippAndMarkusPods}.
We will take a closer look at Model-Faithful Acyclicity (MFA) \cite{MFA}, 
which also gives rise to more specialized notions like Disjunctive MFA (DMFA) \cite{DMFA} and Restricted MFA (RMFA) \cite{RMFA}.
These are some of the most general but also most technical sufficient conditions for chase termination.

\subsection{The Need for Formal Verification}

Especially for chase termination, intuitions often seem simple but start to dissolve when taking a closer look.
Let us consider two interesting examples with subtle but crucial mistakes.
(A) Claim: A particular variant of chase termination is $\Sigma_1^0$-complete (aka. RE-complete) \cite{AnatomyOfTheChase}. 
This problem was shown to be $\Pi_2^0$-hard later \cite[Proposition 42]{NormalisationsOfExistentialRules} and recently proven to be $\Pi_1^1$-complete \cite{ChaseTerminationAnalyticalHierarchy}.
As discussed by the authors, there seems to be an intuitive semi-decision procedure. Only taking a closer look reveals that this does not work because of the ``fairness'' requirement of the chase.
(B) Claim: Restricted Model-Faithful Cyclicity (RMFC) guarantees chase non-termination \cite{RMFA}. 
There is a counterexample for the original proof idea. This was discovered and fixed later when Restricted Prefix Cyclicity (RPC) was introduced as a more refined notion \cite{RPC}.
Similar to MFA-like conditions, RMFC and RPC have suprisingly simple intuitions but extremely technical definitions, which makes catching (subtle) mistakes very difficult.

Besides theoretical research, practical reasoning systems for existential rules and other Datalog-like languages 
have become complex pieces of software with many additional features and performance optimizations that 
go far beyond the theoretical description of the chase \cite{rdfox,vlog,souffle,nemo24}. 
One of the main selling points of these systems is that they are inherently trustworthy because of the conceptual simplicity of rule-based reasoning. 
But with more complex optimizations, more parts need to be trusted.
Formal verification can ensure a high degree of trust but fully verifying state of the art systems is extremely challenging.
However, at least for (stratified) Datalog, 
fully verified systems have been implemented in the Rocq theorem prover~\cite{datalogCoq,datalogCoqThesis,regularDatalog}.
Other works also describe Datalog semantics in various theorem provers to formalize security logics~\cite{certifiedSecturityLogic} (Rocq), 
verify specific Datalog programs used in program analysis~\cite{verifiedProgramAnalysis} (Isabelle/HOL), 
or check certificates provided by a Datalog engine to validate individual reasoning results~\cite{verifyingDatalog} (Lean).
To the best of our knowledge, none of these or other works consider existential rules though.

\subsection{Our Contribution -- A Lean Library}

In this work, we present a formal library of theoretical results around existential rules and the chase, written in Lean~\cite{Lean4}. 
We are inspired and motivated by existing libraries on mathematics\footnote{\url{https://github.com/leanprover-community/mathlib4}} \cite{mathlib}
and (other) computer science topics\footnote{\url{https://github.com/leanprover/cslib}} \cite{cslib} that are actively developed by the Lean community.
Our primary goal is to start a similar effort for theoretical research on existential rules and the chase, 
with the clear benefit of having machine-verifiable and extensible results.
As a fortunate side effect, the mathematical description of the chase on existential rules also lays the foundation 
for the verification of specific implementations and real-world systems. 
For example, a (verified) reference implementation can be built on top of our project, directly in Lean. 

Lean is based on dependant type theory and is a purely functional programming language where propositions are types. 
Proofs are in turn elements of such types. 
That means, the Lean kernel verifies our proofs by type checking.
To prove proposition $Q$ from proposition $P$, we write a function that takes an element of $P$ and transforms it into an element of $Q$. 
The use of ``tactics'' allows us to write proofs in a more human readable format.
Still, the proofs are meant to be explored in an interactive fashion, where a widget in the code editor shows the current hypotheses and the proof goal for the current line.
In this way, one can ``step'' through the proof line by line.
For a comprehensive view on Lean, we refer the interested reader to the official resources\footnote{\url{https://lean-lang.org/learn/}}.

%
%

In the upcoming sections, we present our key ideas behind crafting a Lean library for existential rules and the chase and demonstrate its capabilities
by proving known properties mentioned above. We also generalize and extend some of them. 
Since the code is already thorougly documented, we focus less on the specific implementation but rather on conceptual decisions and insights 
that do not only apply to Lean. 
We phrase main takeaways in \emph{remark} environments.
In the early sections we cover the following questions: 
\begin{itemize}
  \item How do we introduce ``fresh'' terms? (Section~\ref{sec:terms})
  \item When is a rule applicable? (Section~\ref{sec:triggers})
  \item How can we view the chase as an infinite structure but still easily induce over its elements? (Section~\ref{sec:chase})
\end{itemize}
We then show the following results, again using \emph{remarks} to highlight how earlier design decisions come into play.
\begin{itemize}
  \item The chase result is a universal model set (Section~\ref{sec:universality}).
  \item A chase without alternative matches yields a universal model that is a \emph{strong core}, generalizing the original theorem, which uses \emph{weak cores} (Section~\ref{sec:core}). 
  \item We generalize the idea behind (D/R)MFA into a common framework with support for constants in rules and prove that chase termination is guaranteed (Section~\ref{sec:mfa}). 
\end{itemize}
This paper is best read while exploring the formalization in a code editor with a Lean extension installed. This is probably easiest to setup in VSCode.
Also, throughout the code, doc comments provide additional detail on individual results and sometimes larger parts of the formalization. 
Auto-generated documentation is available online at \url{https://monsterkrampe.github.io/Existential-Rules-in-Lean/}.
All of the code (about 19000 lines in total) is available on GitHub where many symbols in this PDF directly link to.
\begin{itemize}
  \item \href{\basicDatastructuresRepoUrl}{github.com/monsterkrampe/Basic-Lean-Datastructures}
  \item \href{\treeRepoUrl}{github.com/monsterkrampe/Possibly-Infinite-Trees}
  \item \href{\ruleRepoUrl}{github.com/monsterkrampe/Existential-Rules-in-Lean}
\end{itemize}

\section{Terms, Atoms, and Rules}\label{sec:terms}

Towards the end of this section, we want to be able to formalize (disjunctive) (existential) rules.
Here, we also discuss our first important design decision of using Skolem terms instead of (labelled) nulls (Remark~\ref{rem:skolemTerms}).
Traditionally, a rule is an expression of the form

$$\forall \vec{x}, \vec{y}. B[\vec{x}, \vec{y}] \to \bigvee_{i=1}^n \exists \vec{z}_i. H_i[\vec{y}_i, \vec{z}_i]$$

where $\vec{x},\vec{y},\vec{z}_1,\dots,\vec{z}_n$ are pairwise-disjoint lists of variables, $\vec{y}_1, \dots, \vec{y}_n$ are (not necessarily disjoint) lists of variables 
such that $\vec{y} = \bigcup_{i=1}^n \vec{y}_i$, and $B$ (body) and $H_1, \dots, H_n$ (head) are function-free conjunctions of atoms.

In Lean, we simplify this into the following structure for \Rule{}s.
Existential variables are identified implicitly: 
we treat variables as existentially quantified if they occur somewhere in \texttt{head}
but not in \texttt{body}.

\begin{lstlisting}
structure Rule where
  id : Nat
  body : FunctionFreeConjunction sig
  head : List (FunctionFreeConjunction sig) 
      -- the List represents a disjunction
\end{lstlisting}

To formalize the basic building blocks, we start with \emph{predicate symbols}, \emph{variables}, and \emph{constants}. 
In Lean, we capture these in a \Signature, usually called \lstinline{sig}. The signature is a structure of three arbitrary types \lstinline{C,V,P} and an arity function \lstinline{P -> Nat}.
As we saw in Example~\ref{exp:mainExample}, we also need to be able to introduce fresh terms. 

\begin{remark}\label{rem:skolemTerms}
  Picking fresh elements from a (countably infinite) set of \emph{(labelled) nulls} is a popular option.
  When ``applying'' a rule, the nulls picked for the existential variables must not occur anywhere else. 
  The problem is that picking is a magical procedure.
  We know that it is safe to do since we have a countably infinite set of nulls at our disposal. 
  But implementing a specific procedure or even only formally stating that a fresh null does not occur ``before'' would require us to keep some global state. 
  Therefore, we use Skolem terms as a naming convention for nulls, which convey only local information and are ``fresh by design''.
\end{remark}

We identify Skolem function symbols (\SkolemFS{})
by a rule, a \lstinline{disjunctIndex}, i.e. the index of the head disjunct in the rule, 
the (existential) variable that gets Skolemized, and the arity of the function symbol.
The function symbols then receive the frontier of the rule as the arguments to construct a Skolem term. 
The \emph{frontier} is the list of variables that occur in both head and body of the rule.
This makes the Skolem terms unique.
This way of Skolemizing with the frontier is also called \emph{semi-oblivious} opposed to \emph{oblivious} Skolemization 
where all body variables would be used.
The semi-oblivious idea is that values used for non-frontier body variables are irrelevant to the result of the rule application.

\begin{lstlisting}
structure SkolemFS (sig : Signature) where
  ruleId : Nat
  disjunctIndex : Nat
  var : sig.V
  arity : Nat
\end{lstlisting}

We define a variety of inductive types to represent different kinds of terms such as \VarOrConst{}, \SkolemTerm{}, and \GroundTerm{}\footnote{The definition for \GroundTerm{}s is a bit more technical but abstracted away to look like an inductive type from the outside.}. 
Instead of having a common term type, this more fine-grained approach limits the number of cases that we need to consider in proofs.
Different kinds of atoms can now be introduced as variants of the following \GeneralizedAtom{} where \lstinline{T} represents an abitrary term type.
\Atom{}s, \FunctionFreeAtom{}s, \Fact{}s, and \FunctionFreeFact{}s are all \GeneralizedAtom{}s using \SkolemTerm{}s, \VarOrConst{}s, \GroundTerm{}s, and \lstinline{sig.C} as their term type, respectively.

\begin{lstlisting}
structure GeneralizedAtom 
    (sig : Signature) (T : Type u) where
  predicate : sig.P
  terms : List T
  arity_ok : terms.length = 
             sig.arity predicate
\end{lstlisting}

A \FunctionFreeConjunction{} is consequently a list of \FunctionFreeAtom{}s, which gives us the essential ingredient for the \Rule{} definition above.
To define a \KnowledgeBase{}, which is going to be the input of the chase, we still require \FactSet{}s, \Database{}s (finite sets of function free facts), and \RuleSet{}s.
These are readily defined with the following basic \Set{} definition:
\lstinline{def Set (α : Type u) := α -> Prop}
That is, a set is a function that takes a potential element and returns whether it is a member.
A \Set{} is \SetFinite{} if there exists a list with the same elements.\footnote{\lstinline{List} is inductively defined and therefore each list is finite.}
For \RuleSet{}s we make the technical demand that rules within the set are uniquely identified by their id, formally: if two rule ids are equal, then so are the rules.

\section{Triggers and Obsolescence}\label{sec:triggers}

As a precursor to the chase, we need to model what it formally means to ``apply'' a rule.
The goal of an application is to satisfy the rule in the most general way.
A major question to ask here is: when should a rule be applied? 
This will dictate which variant of the chase algorithm we model. That is, we could also ask: 
Do we want to model the Skolem or restricted (aka. standard) chase? (Remark~\ref{rem:obsolescence})
Before we answer this question, we cover the basic definitions of \emph{triggers} captured in a structure called \PreTrigger{}.

\begin{lstlisting}
structure PreTrigger (sig : Signature) where
  rule : Rule sig
  subs : GroundSubstitution sig
\end{lstlisting}

\GroundSubstitution{}s are (total) functions from \lstinline{sig.V} to \GroundTerm{}s.
A \PreTrigger{} \lstinline{trg} derives new facts in its \emph{result}, i.e. \lstinline{trg.mapped_head}, by instantiating the Skolemized variant of \lstinline{trg.rule}
with \lstinline{trg.subs}. Thereby, it produces a list of \Fact{}s for each head disjunct.

In the context of any chase variant, a \PreTrigger{} needs to be \loaded{} for a given \FactSet{} $F$. 
That is, \lstinline{trg.mapped_body}, which results from applying \lstinline{trg.subs} to \lstinline{trg.rule.body}, needs to be contained in $F$.
Depending on the chase variant, a trigger can be \emph{obsolete} in different ways. 
The two variants we want to consider here are 
\SkolemObsolescence{} (from the Skolem chase) and 
\RestrictedObsolescence{} (from the restricted chase).
\SkolemObsolescence{} marks a trigger \lstinline{trg} as obsolete for a \FactSet{} $F$, if the result for one head disjunct is already contained in $F$. 
\RestrictedObsolescence{} additionally checks if such a containment can be achieved by allowing a different mapping of the existential variables in \lstinline{trg.rule.head}. 

\begin{example}\label{exp:mainExampleTriggers}
  Reconsidering Example~\ref{exp:mainExample}, 
  we observe that the root node with fact $P(a,b)$ yields two fact sets through the result of the \PreTrigger{} 
  \texttt{trg}
  with the single \Rule{} in $R$ and a \GroundSubstitution{} that maps 
  $x$ to $a$ and $y$ to $b$. The trigger \texttt{trg} is \loaded{} for $\{P(a,b)\}$ since 
  the instantiated body of \texttt{trg.rule} is $P(a,b)$.
  Additionally, \texttt{trg} is obsolete for neither \SkolemObsolescence{} nor \RestrictedObsolescence{}
  since \lstinline{trg.rule.head} is not satisfied by the single fact $P(a,b)$.
  On the other hand, the \PreTrigger{} \texttt{trg'} mapping $x$ to $b$ and $y$ to $f^{1,1}_z(b)$ is \loaded{} 
  but obsolete with respect to \RestrictedObsolescence{} for 
  the fact set $\{ P(a, b), P(b, f^{1,1}_z(b)), P(f^{1,1}_z(b), b) \}$ since its head is satisfied by the facts $P(f^{1,1}_z(b), b)$ and $P(b, f^{1,1}_z(b))$.
  For \SkolemObsolescence{}, \lstinline{trg'} is not obsolete for this fact set but the trigger 
  mapping $x$ to $a$ and $y$ to $b$ would be obsolete for the fact set $\{ P(a,b), S(b) \}$.
\end{example}

Despite the obvious differences of the specific conditions, 
their behavior has certain commonalities that we may extract into a more general class of conditions.

\begin{remark}\label{rem:obsolescence}
  Instead of focusing only on Skolem or restricted obsolescence (and therefore the respective chase),
  we introduce a generalizing \ObsolescenceCondition{} that only demands four essential properties.
  (1) Subset monotonicity: if a trigger is obsolete for a fact set, then it is obsolete on all supersets.
  (2) If a trigger is obsolete, it is satisfied (under first-order logic semantics).
  (3) If the trigger result already occurs in the given fact set, then it is obsolete for the generic condition. 
  (4) If two triggers are equivalent (\PreTriggerEquiv{}), then one is obsolete if and only if the other is obsolete.
  Note that the most liberal condition that still fulfills (3) is \SkolemObsolescence{} and the most restrictive condition that still fulfills (2) is \RestrictedObsolescence{}.
\end{remark}

Now, a \Trigger{} is a \PreTrigger{} with a fixed \ObsolescenceCondition{}. We say that a \Trigger{} is \TriggerActive{} (for a fact set) if it is \loaded{} and not obsolete with respect to its obsolescence condition. 
Finally, an \RTrigger{} is subtype of \Trigger{} enforcing that the trigger's rule stems from a given rule set. For example, 
in a chase for a given knowledge base $K$, it only makes sense to consider triggers that feature rules from $K$ and it is convenient for us
to capture this property directly in the trigger type instead of demanding it explicitely in all places.

Let us briefly mention two other popular chase variants that we do not yet capture by generalizing obsolescence: the oblivious chase and the core chase.
Both differ in ways other than obsolescence. 
To model the oblivious chase, the Skolem terms would need to take all body variables as arguments (not only \lstinline{trg.rule.ruleFrontier}). \SkolemObsolescence{} then gives the oblivious behavior. 
The core chase uses restricted obsolescence but on top of computing trigger results, the resulting fact sets need to be condensed to their ``core'' during the chase. 
In the future, we plan to introduce additional generalizations to be able to capture all four chase variants in a common framework.

\begin{figure*}[t]
\begin{lstlisting}
structure ChaseDerivation (obs : ObsolescenceCondition sig) (rules : RuleSet sig) where
  branch : PossiblyInfiniteList (ChaseNode obs rules)
  -- The derivation needs to contain an initial fact set.
  isSome_head : branch.listHead.isSome 
  -- Each fact set is produced by a trigger.
  triggers_exist : ∀ n : Nat, ∀ before ∈ (branch.drop n).listHead,
    let after := (branch.drop n).listTail.listHead
    (exists_trigger_opt_fs obs rules before after)
  -- The used triggers are active.
  triggers_active : ∀ n : Nat, ∀ before ∈ (branch.drop n).listHead,
    ∀ after ∈ (branch.drop n).listTail.listHead, 
    ∃ orig ∈ after.origin, orig.fst.val.triggerActive before.facts
  -- Every trigger is eventually inactive.
  fairness : ∀ trg : (RTrigger obs rules), ∃ i : Nat, 
    (∃ node ∈ (branch.drop i).listHead, ¬ trg.val.active node.facts)
    ∧ (∀ j : Nat, ∀ node2 ∈ (branch.drop i).listTail.get? j, ¬ trg.val.active node2.facts)
\end{lstlisting}
\caption{Definition of a \ChaseDerivation{} based on \PossiblyInfiniteList{} and four conditions.}\label{listing:chaseDerivation}
\end{figure*}

\section{The Chase as a Coinductive Structure}\label{sec:chase}

As briefly outlined in Example~\ref{exp:mainExample},
the intuitive idea behind the chase is to satisfy all rules in a given \KnowledgeBase{} 
by iteratively and exhaustively adding the results of active triggers to the previously produced \FactSet{}s based on the initial \Database{}.
This procedure yields a possibly infinite tree of finite degree. 
In the special case of \emph{deterministic} rules, where each rule has exactly one disjunct, this tree only consists of a single branch. 
An important design decision will be to take a coinductive view on the chase (Remark~\ref{rem:coinduction}).

\subsection{Chase Derivations and Branches}

Works that are not concerned with disjunctions usually model the chase as an infinite sequence of fact sets.
Even in the disjunctive setting, we can characterize individual branches of a chase tree independently of the tree structure.
For simplicity, we will mainly discuss this view. Lifting the ideas to a whole tree structure is not much harder.
Let us define such a chase branch on paper first, similar to definitions one would usually find in the literature.

\begin{definition}\label{def:chaseBranch}
  For a knowledge base $\langle R, D \rangle$, a \emph{chase branch} is a sequence of fact sets $F_0 = D, F_1, \dots$
  such that
  \begin{itemize}
    \item for each $i \geq 0$, $F_{i+1}$ results from $F_i$ by adding the result of a disjunct of a trigger that is active for $F_i$, and
    \item for each trigger $\lambda$, there is some $F_i$ such that, for each $j \geq i$, $\lambda$ is not active for $F_j$. 
  \end{itemize}
\end{definition}

Let us reflect on what an infinite sequence really is and, more importantly, what it is \emph{not}.
An infinite sequence is \emph{not} an inductive type. Elements of inductive types (like lists) are always finite. 
Think of the natural numbers: a number is either zero or the successor of a number. While there are infinitely many natural numbers, each natural number is finite. 
The elements of an inductive type are the elements of the least fixed point of everything that can be built from the constructors.
What we want instead is the greatest fixed point, i.e. coinductively defined lists.
To model this, we build a small framework around what we call \PossiblyInfiniteList{}.\footnote{Mathlib provides very similar machinery in Stream'.Seq.}
This includes basic access functions such as \ListHead{} and \ListTail{}, a suffix relation (\lstinline{<:+}) and a recursion/induction principle to show properties for elements of the list (\ListMemRec{}).

Based on this insight, we also want to treat the chase branch as a coinductive data structure.
But this does not work with Definition~\ref{def:chaseBranch}.
We currently force chase branches to start on a database, which means that (most) subbranches are no proper chase branches, since their first fact set may already contain Skolem terms.
Therefore, we define a \emph{chase derivation} where we lift this restriction. The other properties are the same as in Definition~\ref{def:chaseBranch}.

\begin{remark}\label{rem:coinduction}
  We treat a \ChaseDerivation{} as a coinductive data structure. 
  From our point of view, this resonates well with proofs about the chase usually carried out on paper only that the formal basics are rarely explicitely discussed.
  For example, when showing properties of fact sets in the chase via induction, it might not always be clear what we induce over. 
  We could use the index within the chase but it is more convenient to consider only the following two cases: the fact set is the initial one or it results from a trigger application.
  This induction principle is something we can define using our coinductive framework (see \ChaseDerivationMemRec).
\end{remark}

Our framework around \ChaseDerivation{}s offers many convenience features inspired by, but also going beyond, the underlying \PossiblyInfiniteList{}.
\begin{itemize}
  \item Accessor functions such as \ChaseDerivationHead{}, \ChaseDerivationNext{}, and \ChaseDerivationTail{}.
  \item A suffix relation on chase derivations (\ChaseDerivationSuffix, where \lstinline{cd2 <:+ cd} reads as: "cd2 is a subderivation of cd").
  \item A recursion/induction principle to show properties over all members of the chase derivation (\ChaseDerivationMemRec).
  \item A total order on derivation members (\ChaseDerivationPrec, where \lstinline{node ≼ node2} reads as: "node occurs in the chase before node2"). A strict version (\lstinline{≺}) also exists.
\end{itemize}

Listing~\ref{listing:chaseDerivation} shows the full \ChaseDerivation{} definition\footnote{In reality, the \ChaseDerivation{} extends an even more general \ChaseDerivationSkeleton{}. Listing~\ref{listing:chaseDerivation} shows the combined structure.}. Note the similarities to Definition~\ref{def:chaseBranch}.
The associated conditions are still expressed in the vocabulary of the underlying \PossiblyInfiniteList{}, which leaves them quite convoluted. 
With the new vocabulary, we show theorems that express the conditions in a more accessible way. 
For example, we restate fairness in \ChaseDerivationFairnessPrime{} and \ChaseDerivationFairnessPrec{} in terms of suffixes and the predecessor relation, respectively.
One can see from the definition that the \ChaseDerivation{} is generic over an \ObsolescenceCondition{}, which was introduced in the last section. 
Furthermore, maybe unexpectedly, the derivation is not based on a possibly infinite list of \FactSet{}s but \ChaseNode{}s instead. 
A \ChaseNode{} stores a \FactSet{} and an optional \RTrigger{}. So each chase step remembers not only \emph{what} was derived but also \emph{how}.
A \ChaseBranch{} as in Definition~\ref{def:chaseBranch} is a \ChaseDerivation{} for a \KnowledgeBase{} where the first fact set is forced to be the \Database{}.

Let us now discuss some important properties about \ChaseDerivation{}s and \ChaseBranch{}es. 
We already mentioned that we define a total order on the \ChaseNode{}s of the derivation. But so far it might not be obvious that this is possible.
Could the same \ChaseNode{} not occur multiple times in the same derivation? Fortunately, no!
The essential theorem for this shows that the \ChaseDerivationHead{} can never also be a member of the \ChaseDerivationTail{}. 
Assuming for a contradition that this was the case, the copy of head was introduced by a trigger, but then the trigger cannot be active anymore since all the derived facts already occur in the original head. We rely on property (3) from Remark~\ref{rem:obsolescence} here.

Furthermore, we mentioned in Remark~\ref{rem:skolemTerms} that Skolem terms are ``fresh by design''. We now want to prove this along a chase branch.
What we show is slightly different but entails the desired property: if a Skolem term for a trigger $\lambda$ occurs in a \ChaseNode{}, then the whole result of $\lambda$ is already in this chase node.
This means if a Skolem term is already present, then every trigger that would introduce it (again) is already obsolete, which ensures freshness.
Note that this is not quite the case for derivations instead of branches, since the term in question could occur in the first fact set instead of resulting from a trigger.

The most important property of a chase branch is that its result, defined as the union of all fact sets, is a model (\ChaseBranchResultModelsKb{}).
Without disjunctions, one obtains a stronger result showing that the result of a chase branch on deterministic rules is a \emph{universal model}, meaning that it is one of the most general ones 
(\DeterministicChaseBranchResultUniversallyModelsKb{}). Our proof for this depends on a corresponding result on chase trees, which we discuss at length in Section~\ref{sec:universality}.

%

\begin{example}\label{exp:mainExampleBranch}
  Reconsidering Example~\ref{exp:mainExample} 
  and our insights from Example~\ref{exp:mainExampleTriggers},
  we can construct a finite \ChaseBranch{} for \RestrictedObsolescence{} 
  that contains exactly $P(a,b), \allowbreak P(f^{1,1}_z(b), b), \allowbreak P(b, f^{1,1}_z(b))$ in its result. 
  We can also construct an infinite \ChaseBranch{} for \SkolemObsolescence{}
  that derives an infinite ``chain'' of $P$ relations including 
  their inverses except for $P(b, a)$.
  Note that both of these results are models of the input knowledge base. 
  Furthermore, there is a \ChaseBranch{} producing only $P(a,b), S(b)$ for both 
  obsolescence conditions, which also models the knowledge base.
\end{example}

\subsection{Tree Derivations and Chase Trees}

\TreeDerivation{}s and \ChaseDerivation{}s are similar. We mainly exchange the \PossiblyInfiniteList{} for a \FiniteDegreeTree{}.
We provide a similar coinductive framework to access the root, the (immediate) child nodes and child trees, as well as a subtree relation and a recursion/induction principle for members.
The only drawback we have is the predecessor relation. 
Instead of defining the relation directly on the \ChaseNode{}s, 
we now need to take their addresses in the tree into account to obtain a proper order.
This is because the same node can now occur multiple times in the tree, just not along the same branch.
The address is an easy way to tell the different occurrences apart.
The predecessor relation is then merely the prefix relation of the addresses.
Just like the \ChaseDerivation{}, the \TreeDerivation{} does not enforce starting on a \Database{}.
In a similar way that \ChaseBranch{}es are defined for \ChaseDerivation{}, we also define \ChaseTree{}s based on \TreeDerivation{}s.

Each branch in tree of the \TreeDerivation{} is shown to be a proper \ChaseDerivation{} and 
the set of branches for the \TreeDerivation{} consists of all \ChaseDerivation{} that correspond to a branch in the tree of the \TreeDerivation{}.
We can now define the \TreeDerivation{} result as follows. 

\begin{lstlisting}
def treeDerivationResult (td : TreeDerivation obs rules) : 
    Set (FactSet sig) := 
  td.branches.map ChaseDerivation.result
\end{lstlisting}

For the special case of a \ChaseTree{}, it is easy to show that every fact set in the result is a model of the underlying knowledge base
since we already know that the result of each \ChaseBranch{} is a model (see \ChaseTreeResultModelsKb).

\begin{example}\label{exp:mainExampleTree}
  Based on Example~\ref{exp:mainExampleBranch},
  we can see that there is a (single) \ChaseTree{} 
  for \RestrictedObsolescence{} and the respective knowledge base. 
  This tree has a root node with $P(a,b)$ and two children 
  adding $S(b)$ or $P(f^{1,1}_z(b), b), \allowbreak P(b, f^{1,1}_z(b))$, respectively.
  The result therefore is the set with the two \FactSet{}s: 
  $\{ P(a,b), S(b) \}$ and $\{ P(a,b), \allowbreak P(f^{1,1}_z(b), b), \allowbreak P(b, f^{1,1}_z(b)) \}$.
  We already realized earlier that both are models of the knowledge base.
\end{example}

The next section focuses entirely on showing a universality property for the \ChaseTree{} result.

\begin{figure*}[t]
\begin{lstlisting}
abbrev InductiveHomomorphismResult (ct : ChaseTree obs kb) (m : FactSet sig) :=
  { pair : ct.NodeWithAddress × (GroundTermMapping sig) // 
    pair.snd.isHomomorphism pair.fst.node.facts m }

noncomputable def hom_step (m_is_model : m.modelsKb kb)
  (prev_res : InductiveHomomorphismResult ct m) : Option (InductiveHomomorphismResult ct m)
\end{lstlisting}
\caption{Signature of Inductive Homomorphism Definition for the Universality Proof.}\label{listing:inductiveHomomorphism}
\end{figure*}

\section{Chase Trees yield Universal Model Sets}\label{sec:universality}

The universality property of a \ChaseTree{} is as follows.
A set of models $\mathcal{U}$ is a \emph{universal model set} if for each model $M$, there is a model $U \in \mathcal{U}$ such that
there is a homomorphism from $U$ to $M$.
So we aim to show that the \ChaseTree{} result is a universal model set.

\begin{lstlisting}
theorem chaseTreeResultIsUniversal 
    (ct : ChaseTree obs kb) : 
  ∀ (m : FactSet sig), m.modelsKb kb -> 
    ∃ (fs : FactSet sig) 
    (h : GroundTermMapping sig), 
    fs ∈ ct.treeDerivationResult ∧ h.isHomomorphism fs m
\end{lstlisting}

\begin{example}\label{exp:mainExampleUniversality}
  Recall the result from the \ChaseTree{} discussed in Example~\ref{exp:mainExampleTree},
  that is the set $\mathcal{U}$ with the two fact sets $\{ P(a,b), S(b) \}$ and $\{ P(a,b), \allowbreak P(f^{1,1}_z(b), b), \allowbreak P(b, f^{1,1}_z(b)) \}$.
  We want to argue intuitively that $\mathcal{U}$ is universal.
  Every model for the input knowledge base needs to contain $P(a,b)$ and satisfy the single rule 
  from Example~\ref{exp:mainExample}. 
  But for this to be true, it needs to contain $S(b)$
  or it needs to contain $P(b,n), P(n,b)$ for some term $n$.
  For any such model $M$, we find a \GroundTermMapping{} $h$, which is the identity on all terms 
  except that it maps $f^{1,1}_z(b)$ to $n$.
  Thereby, we can pick either the first or second element from $\mathcal{U}$ 
  (call this $U$) and witness that 
  $h$ is a homomorphism from $U$ to $M$.
\end{example}

A proof on paper might just say that for a given $M$, the corresponding $U$ and the homomorphism can be constructed inductively along the chase.
While this is an essential ingredient, this does not conclude the proof. The final $U$ only results from taking the union of the constructed sets 
and the homomorphisms also still need to be combined into a single function. And even then, we still need to prove that the combined $U$ indeed occurs in the 
result of the chase tree, which is to 
specify the infinite branch that the constructed fact sets belong to.
And of course, we also need to prove that the combined homomorphism agrees with the 
inductively constructed ones on their corresponding fact sets.
All of this is not hard to believe and therefore likely omitted on paper. 
Still the formal arguments are not trivial and yet again involve some coinductive principles.

\begin{remark}\label{rem:generateBranch}
  To be able to construct an infinite branch in a tree from an inductive construction, 
  we make use of the \GenerateBranch{} function, which is part of our \FiniteDegreeTree{} framework.
  This function is \emph{corecursive} and accepts a ``generator'' function as input that always creates a new tree element from a previous one.
  Given that this generator adheres to certain properties, we can directly conclude that the \PossiblyInfiniteList{} produced by the 
  \GenerateBranch{} function is indeed a branch in the \FiniteDegreeTree{}.
  We also lift this to \TreeDerivation{}s in \TreeDerivationGenerateSubderivation{}.
\end{remark}


The generator function (\HomStep{}) takes an \InductiveHomomorphismResult{} as input and output, which consists of a node in the chase tree and a \GroundTermMapping{} that is a homomorphism from the node into the target model $M$.
We show the type signatures in Listing~\ref{listing:inductiveHomomorphism}.
The application of \TreeDerivationGenerateSubderivation{} can be found in the proof of \ChaseTreeResultIsUniversal{}.
After setting up an initial \InductiveHomomorphismResult{}, we essentially call 
\TreeDerivationGenerateSubderivation{} passing 
\lstinline{(hom_step ct m m_is_model)} as the generator
to obtain the full branch that yields $U$ as its result.
Using \TreeDerivationGenerateSubderivationMemBranches{}, we get the proof that the generated branch indeed occurs in the branches of the chase tree.
We also define a combined version of all homomorphisms (called \lstinline{global_h}) and show all necessary properties, which involves quite a few lines of code, but this is conceptually less interesting.
Thus, we have now successfully shown in Lean that every \ChaseTree{} for any \ObsolescenceCondition{} indeed procudes a universal model set, which, as mentioned earlier,
also yields the special case: when restricting to deterministic rules, the result of every chase branch for any \ObsolescenceCondition{} is a universal model.

\section{Alternative Matches and Cores}\label{sec:core}

In this section, we discuss that, going beyond universality, if 
we can avoid \emph{alternative matches} in a \ChaseBranch{} on deterministic rules, then the result is a core.
In fact, we have formalized all results from Section~3 in \cite{RestrictedChaseCores}.
We even generalize the presented results as we again show the results generically for any \ObsolescenceCondition{} 
but more importantly, since we consider two different notions of cores.
On infinite structures, different definitions of cores are possible. For clarity, we refer to the definition used in \cite{RestrictedChaseCores} as a ``weak core'', 
which states that a \FactSet{} is a \emph{weak core} if every endomorphism is an embedding (strong and injective).
A \FactSet{} is a \emph{strong core} if every endomorphism is an automorphism (strong, injective, and surjective). 
On finite structures the notions of weak and strong core coincide.
The main goal of this section is to describe the formalization of the following result (which includes a corresponding result for weak cores).

\begin{lstlisting}
theorem result_isStrongCore_of_noAltMatch 
  {cb : ChaseBranch obs kb} 
  (det : kb.knowledgeBaseIsDeterministic) : 
  ¬ cb.has_alt_match -> 
    cb.result.isStrongCore
\end{lstlisting}

Since homomorphisms are \GroundTermMapping{}s and therefore total functions, defined on all terms, our 
definitions of \lstinline{strong!} and \lstinline{injective_for_domain_set} take an argument that specifies 
the relevant part of the domain. 
For example, 
for the definition of \lstinline{strong!}, one might intuitively state 
that the mapping of every fact not in a given fact set \lstinline{fs} should again not be in \lstinline{fs}. 
This demand is too strong. Often, we have no information on how a homomorphism handles terms that are outside of \lstinline{fs.terms}
and there might very well be facts outside of \lstinline{fs} that end up in \lstinline{fs} under the homomorphism. 
As long as such a fact features at least one term outside of \lstinline{fs.terms}, this is unproblematic.
The requirement for strong homomorphisms shall only ensure that previously existing but unconnected terms do not get connected via a predicate by the homomorphism.

\begin{lstlisting}
def strong! (h : GroundTermMapping sig) 
    (domain : Set (GroundTerm sig)) 
    (A B : FactSet sig) : Prop :=
  ∀ (e : Fact sig), 
    (∀ t, t ∈ e.terms -> t ∈ domain) -> 
    ¬ e ∈ A -> ¬ (h.applyFact e) ∈ B
\end{lstlisting}

\begin{figure*}[t]
\begin{lstlisting}
def isAlternativeMatch (h_alt : GroundTermMapping sig) (trg : PreTrigger sig) 
    (disj_index : Fin trg.mapped_head.length) (fs : FactSet sig) : Prop :=
  (h_alt.isHomomorphism trg.mapped_head[disj_index.val].toSet fs) ∧
  (∀ t, t ∈ trg.rule.ruleFrontier.map trg.subs -> h_alt t = t) ∧
  (∃ t, (t ∈ trg.fresh_terms_for_head_disjunct disj_index.val ...) ∧
        (¬ t ∈ (trg.fresh_terms_for_head_disjunct disj_index.val ...).map h_alt))
\end{lstlisting}
\caption{Definition of Alternative Match.}\label{listing:alternativeMatch}
\end{figure*}

The idea for alternative matches is to witness redundancies in a \ChaseBranch{}. 
Intuitively, a term $t$ that is newly introduced by a trigger is redundant if we can find a homomorphism from the result of the trigger into 
another part of the chase such that $t$ does not occur in the image.
The existence of an alternative match for a trigger witnesses that at least one newly introduced term is redundant in the context of a given fact set.
We present the formal definition in Listing~\ref{listing:alternativeMatch}. 
We say that a \ChaseBranch{} has an alternative match if some trigger in the branch has an alternative match in the context of the result of the branch.
This definition does not yet make the assumption of deterministic rules. We only demand this on theorems where it is really necessary.

\begin{example}
  Consider the database $D = \{P(a, b)\}$ and the singleton rule set $R$:
  $$P(x, y) \to \exists z. P(y, z) \land P(y, x)$$
  The rule applied to $P(a,b)$ produces $P(b, f^{1,1}_z(b)), P(b, a)$ and thereby 
  a redundancy in itself ($f^{1,1}_z(b) \mapsto a$). 
  Indeed, this mapping witnesses an alternative match.
  Still, $\{ P(a,b), P(b,a) \}$ is a universal model and also a core. Only we cannot obtain it with the restricted or Skolem chase.
\end{example}

For the main goal, we first prove that the chase result is a weak core in the absence of alternative matches in \resultIsWeakCoreOfNoAltMatch{}. 
We follow along the original proof \cite[Theorem 2]{RestrictedChaseCores}.
We assume for a contradiction that some endomorphism on the \ChaseBranch{} result is not strong or not injective. 
However, since the branch does not have alternative matches, we can inductively modify this endomorhpisms to 
be the identity on each step in \ChaseBranch{}.
At the same time, we can show alongside the same induction that this 
construction preserves that the endomorphism is not strong or not injective, respectively, which eventually leads to a contradiction.

\begin{remark}\label{rem:homomorphismExtension}
  For a key lemma, we again utilize a corecursive construction similar to Section~\ref{sec:universality} although a bit simpler. 
  Whenever there is a homomorphism $h$ from a fact set $F$ in a \ChaseBranch{} to its result,  
  we need to be able to extend $h$ into an endomorphism on the result that agrees with the original homomorphisms on $F$.
  We formalize this in \homForNodeExtendableToResult{}
  with an inductive construction of a homomorphism (without constructing a tree branch now). 
  We use the \Generate{} function of \PossiblyInfiniteList{}s to obtain an infinite list of homomorphisms that we combine into one.
\end{remark}

To prove \resultIsStrongCoreOfNoAltMatch, we still need to show that every endomorphism on the chase result $U$ is surjective.
But this is close to the alternative match definition. Suppose for a contradiction that 
some endomorphism $h$ on $U$ is not surjective. Then, there is a first term $t$ that is 
never mapped to itself by $h^i$ for any $i \geq 1$ (indicating that $h$ is repeated $i$ times).
The term $t$ originates from some trigger $\lambda$.
Each term $s$ that was introduced earlier is therefore mapped to itself by $h^j$ for some $j \geq 1$. By picking a large enough common multiple $k$ 
of these $j$, (1) the mapping $h^k$ is still an endomorphism on $U$ and therefore also a homomorphism from the result of $\lambda$ into $U$,
(2) all frontier terms (which occurred before $\lambda$) are mapped to themselves, and 
(3) $t$ does not occur in the image of the fresh terms of the trigger after applying $h^k$.
This shows all three conditions from the alternative match definition and therefore yields the desired contradiction.

\section{A Framework for MFA-like Conditions}\label{sec:mfa}

Model-Faithful Acyclicity (MFA) \cite{MFA}, Disjunctive MFA (DMFA) \cite{DMFA}, and Restricted MFA (RMFA) \cite{RMFA} 
are sufficient conditions for chase termination (aka. acyclicity notions). 
So if a rule set fulfills the condition, then every chase on it terminates.
Termination formally means that the underlying \PossiblyInfiniteList{} is finite, in case of a \ChaseBranch{}, or that all branches are finite, in case of a \ChaseTree{}.
A \KnowledgeBase{} terminates if all of its \ChaseTree{}s terminate
and a \RuleSet{} terminates if all \KnowledgeBase{}s featuring this rule set terminate.
We generalize MFA, DMFA, and RMFA into a common framework for any \ObsolescenceCondition{}
and (as per our usual definition) we allow rules to contain constants, which is not considered in the original definitions of DMFA and RMFA. 
For constants, not many changes are necessary but the challenge is to make the \emph{right changes} in the \emph{right places}.
We highlight these in \textbf{bold}. 
Thanks to Lean, we can easily spot how the changes affect our proofs.
We should also mention that the original definitions for DMFA and RMFA 
are optimized to work only with chase trees that prioritize the application of Datalog rules. 
We drop this optimization in our formalization to support arbitrary chase trees.

The central idea of all 3 acyclicity notions is to compute a chase-like procedure starting on a special database based on a given rule set \lstinline{rs}. 
We call the resulting fact set \mfaSet{} and show that \lstinline{rs.ruleSetTerminates} if \mfaSet{} is finite (see \TerminatesOfMfaSetFinite{}).
The idea for the proof is that every trigger used in any \ChaseBranch{} has a corresponding trigger application in the \mfaSet{}.

The \mfaSet{} is based on the \mfaKb{}, which can be obtained solely from \lstinline{rs}.
The rule set in \mfaKb{} is exactly \lstinline{rs}.
The database of \mfaKb{} (aka. \criticalInstance{}) contains facts for each predicate in \lstinline{rs} and all possible combinations of \textbf{constants from \lstinline{rs}} and a special constant $\star$. 

The \parallelDeterminizedChase{} is a chase-like procedure used to compute \mfaSet{} from \mfaKb{}.
It treats disjunctions as conjunctions and applies all active triggers at once.
It thereby yields an \InfiniteList{} of \FactSet{}s. 
Indeed, this list is always infinite, since when no triggers are active anymore, then the last fact set is just repeated forever.
The goal is that \mfaSet{} is a gross overestimation of all fact sets 
that may occur in any \ChaseBranch{} of \mfaKb{}. At the same time, \mfaKb{} generalizes all \KnowledgeBase{}s featuring \lstinline{rs}.
So the \mfaSet{} really overapproximates the result of any \ChaseBranch{} of any such knowledge base.
The proof of \TerminatesOfMfaSetFinite{} comes down to showing that every \ChaseBranch{} for every \KnowledgeBase{} featuring \lstinline{rs} 
can be embedded into \mfaSet{}. This is done by mapping all constants to $\star$ \textbf{except the ones occurring in the rule set, which are just mapped to themselves}.
We prove this via induction using \ChaseDerivationMemRec{}.

\begin{remark}\label{rem:mfa}
  The realization of Remark~\ref{rem:obsolescence}, namely that we can generalize Skolem and restricted obsolescence into a common \ObsolescenceCondition{},
  resonates well with DMFA and RMFA and immediately allows us to generalize these notions accordingly. 
  Not only that but also MFA can be expressed in a similar fashion.
\end{remark}

The difference between MFA, DMFA, and RMFA comes down to which triggers we consider to be active in the \parallelDeterminizedChase{}. 
We use an \MfaObsolescenceCondition{} (MOC) to express this, which is an alias for \LaxObsolescenceCondition{}. These enforce fewer conditions than our usual \ObsolescenceCondition{} (OC). 
We then make our main result \TerminatesOfMfaSetFinite{} generic over an MOC $m$ and an OC $o$ and demand that $m$ \blocksObs{} $o$. 
The idea behind \blocksObs{} is 
that whenever a trigger $\lambda$ from the \parallelDeterminizedChase{} is obsolete according to $m$, 
then all \emph{corresponding} triggers $\lambda'$ from any real \ChaseBranch{} on any knowledge base are obsolete according to $o$.
Corresponding means that replacing all constants in $\lambda'$ by $\star$ \textbf{except the ones from the rule set} 
yields $\lambda$. Or in other words: $\lambda'$ is simulated by $\lambda$ in the \parallelDeterminizedChase{}.

MFA uses \DeterministicSkolemObsolescence{} (DSO) for its MOC, which marks a trigger as obsolete if all of its results are already present.
For DMFA and RMFA (and everything in between), we use \BlockingObsolescence{} (BO), which depends on a given OC.
For both DSO and BO, we need to show that they have the necessary \blocksObs{} property.
DSO has the \blocksObs{} property for every OC and BO has this property exactly for the OC that it receives as an argument.
So DMFA and RMFA correspond exactly to BO with \SkolemObsolescence{} and \RestrictedObsolescence{} respectively, but any other OC could be used just as well.

The definition of \BlockingObsolescence{} is extremely involved and beyond what we can cover in this writeup. 
To at least give an intuition, the idea for both DMFA and RMFA is to limit the number of triggers in the computation of the \mfaSet{} by 
checking which triggers $\lambda$ are necessarily obsolete as soon as they are loaded. 
The basis for this check is a backtracking of triggers that need to be applied in order for $\lambda$ to become loaded.
This roughly works as follows.
When $\lambda$ is loaded, its instantiated body needs to be present. But not only that. Each Skolem term stems from a unique rule head disjunct. 
Consequently a suitable trigger must have been applied before. 
In this way, one can backtrack a set of facts $F$ that are necessarily involved in the target trigger $\lambda$ becoming loaded. 
If $\lambda$ is obsolete with respect to this $F$, then we call $\lambda$ ``blocked'' since it can under no circumstances be applied in a chase. 
This is exactly when \BlockingObsolescence{} marks $\lambda$ as obsolete.\footnote{The Datalog optimization in the original definitions of DMFA and RMFA computes the Datalog-closure on the backtracked facts.}
Checking whether $\lambda$ is obsolete with respect to $F$ uses the OC that BO received as an argument. The difference in the OC is also precisely the what sets DMFA and RMFA apart.

While the backtracking idea is still intuitive, 
in truth this whole procedure involves introducing fresh constants for terms introduced in the backtracking and renaming constants in triggers apart. 
Fresh constants are necessary because our Skolem terms do not indicate the mapping of non-frontier body variables. 
Picking fresh constants introduces a problem similar to Remark~\ref{rem:skolemTerms} and 
in this case indeed involves passing around a list of constants that have already been used.
Renaming constants in triggers apart is necessary since the BO check 
considers triggers from the \parallelDeterminizedChase{}, 
which would only feature $\star$ \textbf{and the rule set constants} in all positions, 
but the check still needs to generalize all possible triggers that end up mapping to $\lambda$.
Renaming again leads to the introduction of fresh constants.
Specifically the handling of fresh constants is what makes the machinery around \BlockingObsolescence{} involved.
Defining \BlockingObsolescence{} and proving the \blocksObs{} property involves a couple of thousand lines of code.
For more explanations and examples for some of the subtleties, we refer the interested reader mainly to the original papers \cite{DMFA,RMFA}.

There is still one point that we have not discussed. How do we know if \mfaSet{} is finite? The acyclicity notions should be decidable after all and not run into an infinite computation.
All notions handle this in the same way: the computation of the \parallelDeterminizedChase{} stops as soon as we witness a \emph{cyclic term}, e.g. $f(g(f(c), d))$ but not $g(f(c), f(d))$.
Since the function symbols are Skolem functions, a cyclic term hints at a rule being applied on its own result. This does not necessarily endanger termination but it could.
In this case, we cannot tell if \mfaSet{} is finite. 
If the procedure terminates on its own without seeing a cyclic term, then \mfaSet{} is obviously finite.
Since there are only finitely many non-cyclic terms, the procedure always halts.
This is how we define \isMfa{} (again generically over an MOC).
The main correctness result is \TerminatesOfIsMfa{}, which we also instantiate specifically for DSO and BO afterwards.
Convincingly presenting this argument to the Lean kernel involves theorems around finiteness of sets containing different syntactic entities.
For example, we require a result showing that a \FactSet{} must be finite if it only features finitely many different predicates and only finitely many different terms.
Also, we show that a functional term of a large enough depth is necessarily cyclic if there are only finitely many different function symbols
(see \cyclicOfDepthTooBig{}).

\begin{example}\label{exp:mainExampleMfa}
  We realized in Example~\ref{exp:mainExampleBranch}
  that there is an infinite \ChaseBranch{} for \SkolemObsolescence{}. 
  Consequently, the rule set should not be (D)MFA. 
  Indeed, the \mfaSet{} is infinite and 
  contains
  $P(\star,\star), \allowbreak S(\star), \allowbreak P(f^{1,1}_z(\star), \star), \allowbreak P(\star, f^{1,1}_z(\star)), \allowbreak 
  P(f^{1,1}_z(f^{1,1}_z(\star)), f^{1,1}_z(\star)), \allowbreak P(f^{1,1}_z(\star), f^{1,1}_z(f^{1,1}_z(\star)))$. 
  Thus, we witness a cyclic term $f^{1,1}_z(f^{1,1}_z(\star))$ and conclude that the rule set is not (D)MFA.
  However, the rule set is still RMFA.
\end{example}

\section{Concluding Remarks}

We provide a formalization of disjunctive existential rules and a generalized chase procedure in Lean
also modelling possibly infinite trees as a by-product.
We demonstrate the usability 
by proving and extending known results from existential rule research:
(1) the chase result is a universal model set, 
(2) a chase without alternative matches yields a universal model that is a \emph{strong} core, and
(3) we prove correctness of a novel framework built around MFA-like acyclicity notions with full support for constants in rules.

The possibilities for extending this library are just as broad as the realm of existential rule research. 
On our roadmap, addressing issue (B) from the introduction is one of our main upcoming goals. 
The considerations around DMFA and RMFA feel like a mere precursor to this as non-termination conditions such as 
DMFC~\cite{DMFA} and RPC~\cite{RPC} only get more complicated. 
Futhermore, formalizing the core chase is already work in progress. 
Specifically, we aim to show that the core chase terminates if and only if a finite universal model exists \cite[Theorem 7]{ChaseRevisited}.
Related to cores, extending the formalization around alternative matches to cover more parts of the introducing paper is also well within our scope \cite{RestrictedChaseCores}.
Long term, we hope to add the result related to issue (A)
and decidability results for chase termination on existential rule fragments~\cite{LinearSingleHeadTermination,GuardedSingleHeadTermination,LinearMultiHeadTermination},
which require more foundational work on computability theory.

Beyond theoretical results, it would be interesting to have a formally verified reference implementation of the chase or the MFA-like conditions directly in Lean. 
Right now, we only describe the chase structures mathematically without specific procedures for how they can be computed. 
However, this would be possible with a significant but manageable amount of work.
Lean is in principle able to compile computable definitions directly into executable code. 
The biggest issue here likely will be that the chase is not guaranteed to terminate. Lean usually expects recursion to be well-founded. 
If this is not obvious, Lean also expects a termination proof. 
There are ways to define possibly non-terminating functions though, but they come at a cost. 
Definitions can be marked as \lstinline{partial}, which still allows to execute them but we essentially lose the ability to prove properties about them.
What seems more interesting is the rather recent \lstinline{partial_fixpoint} feature. 
It allows to mark certain functions, e.g. tail-recursive ones, as possibly non-terminating. Compared to the \lstinline{partial} keyword, 
we retain the possibility of proving properties about the function.

On a meta level, we hope our humble contribution to be a spark that ignites interest within the community to formalize more 
definitions and results from the field of knowledge respresentation and reasoning.
We believe that a formal library of definitions and proofs not only ensures reproducibility
but also encourages development of novel, fully verified theorems. 
And, as we demonstrate, inspires new generalizations.
We realized in our own formalization efforts that the process can be tedious and time-consuming at times
but pays off and even becomes enjoyable in the long run.

\section*{Acknowledgments}
I want to thank my colleague Dr. Stephan Mennicke for proofreading and providing feedback on a draft of this paper.

This work was partly supported
by DFG (German Research Foundation) in project 389792660 (TRR 248, \href{https://www.perspicuous-computing.science/}{Center for Perspicuous Systems}) 
and in the CeTI Cluster of Excellence;
by the Bundesministerium für Forschung, Technologie und Raumfahrt (BMFTR) 
in the \href{https://www.scads.de}{Center for Scalable Data Analytics and Artificial Intelligence} (ScaDS.AI);
and by BMFTR and DAAD (German Academic Exchange Service) in project 57616814 (\href{https://secai.org/}{SECAI}, \href{https://secai.org/}{School of Embedded and Composite AI}).

\section*{AI Declaration}
The authors have employed Generative AI tools only for proofreading (spellchecking and improvement suggestions).

\appendix

\bibliographystyle{kr}
\bibliography{references}

\end{document}